# Technical Upgrades to and Enhancements of a System Vulnerability Analysis Tool Based on the Blackboard Architecture

Matthew Tassava, Cameron Kolodjski, Jeremy Straub
Institute for Cyber Security Education and Research
North Dakota State University
1320 Albrecht Blvd., Room 258
Fargo, ND 58108
Phone: +1-701-231-8196
Fax: +1-701-231-8255
Email: matthew.tassava@ndsu.edu, cameron.kolodjski@ndsu.edu, jeremy.straub@ndsu.edu

**Abstract**

A system vulnerability analysis technique (SVAT) for the analysis of complex mission critical systems (CMCS) that cannot be taken offline or subjected to the risks posed by traditional penetration testing was previously developed. This system uses path-based analysis of vulnerabilities to identify potential threats to system security. Generalization logic building on the Blackboard Architecture's rule-fact paradigm was implemented in this system, the software for operation and network attack results review (SONARR). This paper presents an overview of additional functionality that has been added to this tool and the experimentation that was conducted to analyze their efficacy and the performance benefits of the new in-memory processing capabilities of the SONARR algorithm. The results of the performance tests and their relation to networks' architecture are discussed. The paper concludes with a discussion of avenues of future work, including the implementation of multithreading, additional analysis metrics like confidentiality, integrity, and availability, and improved heuristic development.

## 1. Introduction

The software for operations and network attack results review (SONARR) is a Blackboard Architecture-based system vulnerability analysis tool (SVAT) introduced in [1]. The SONARR algorithm exhaustively traverses a given model network, employing a knowledge base of potential attacks, to evaluate the security of the network's nodes. While the algorithm traverses the network, it generates attack paths that record the fact alterations performed by the rules in the knowledge base. These paths are plausible routes through the network that attackers can take, based on exploitable vulnerabilities within the network.

This paper introduces numerous changes and enhancements to the SONARR technology. These include the addition of normal facts that do not require an associated common property and can act as environment variables for a network as a whole. Second, normal rules, that can influence facts both inside and outside of the container-link-container connections on which generic rules operate, have been added. Action objects, introduced in [2], have been added to SONARR to allow for command execution on the host operating system. Containers and links have been altered to be entities. This allows links to have facts and to be variants (i.e., objects changed from their original value as part of a path). Active variant entities for reality paths were also added. They provide the most recent changes made to the network within its reality. Rule success and link traversability values have been added as additional object data and to allow rule ordering before traversal. Condition objects were also added to provide rule flexibility and support future scalability.

Additionally, the mechanisms that control traversal processing have been changed. The link cap has been replaced with a path termination heuristic, which terminates a path if the current connection (with its resulting fact-rule evaluation) has already been created within that path. A rule-running heuristic has also been added that limits the number of generic rules run on each connection.

The most noticeable change to SONARR is an algorithm redesign to utilize in-memory objects during traversal processing, instead of string parsing. This change significantly speeds up traversal time; however, it requires a larger amount of memory, which is particularly noticeable for larger network models. This paper provides an overview of these changes and evaluates, in particular, the performance of the new path termination heuristic and its behavior with networks that grow in scale and complexity.

## 2. Background

The Blackboard Architecture is a problem-solving system introduced by Hayes-Roth in [3]. It builds on the concept of an expert system [4] and utilizes rules to assess and alter a collection of facts. The Blackboard Architecture is based on the HEARSAY-II system, that was developed for a DARPA-sponsored competition for speech recognition software [5]. It has been utilized in numerous application areas such as agent coordination in real-time strategy games [6], robotics development [7-10], PROLOG programming [11], medical imaging [12,13], counter-terrorism [14], and software testing [15]. Prior work has expanded upon the Blackboard Architecture with an external operating software command execution mechanism [2]. Prior work has also been conducted regarding the modeling of attack paths with frameworks like MITRE's ATT&CK and Lockheed Martin's Cyber Kill Chain [16].

A key element of the Blackboard Architecture, and expert systems in general, is the flexibility of the logic assessment mechanism. The ability to add and remove knowledge as needed provides system development opportunities such as self-learning [17,18] and scalable distribution computation [19,20]. Expert systems are also readily understandable in their reasoning and explainable to a much greater extent than is typical for algorithms employing machine learning techniques.

Many recent artificial intelligence techniques, such as neural networks and machine learning [21,22] are inherently opaque. The algorithms are complex and their decision-reasoning can be difficult to understand. Therefore, developing eXplainable AI (XAI) mechanisms is an important area of research within the field. Many modern AI systems employing deep learning and similar algorithms can be accurate and consistent in their output. However, without their processing being human-understandable, it is difficult to trust the conclusions drawn [23,24]. Even more readily understandable mechanisms, like a knowledge-based architecture, still present a challenge because explainable AI systems need to clarify how they arrived at a conclusion, when it is challenged or misunderstood [25].

One area that explainable AI can contribute significantly to is automated tools for cybersecurity purposes. As technology and consumer demand grows so does the complexity of networks' services [26]. This makes it increasingly difficult for cybersecurity professionals to monitor the network infrastructure, identify vulnerabilities, and respond to active security threats. Penetration testing is regarded as an important activity to secure a network [27]. Network complexity, though, makes human penetration difficult and increases the likelihood of vulnerabilities being missed. Development of tools that replicate the human perceptive element and learn to adapt to new attacks is a key area of research in this field, and one that explainable AI holds potential promise for. Automation tools and frameworks already exist for various other cybersecurity tasks like log analysis [28-30], dynamic malware analysis [31], and intrusion detection [32,33].

## 3. System Overview

This section provides an overview of the SONARR system and its implementation. SONARR is comprised of three application interfaces: SONARR, ADM, and TRDM. The SONARR interface performs a network traversal of an imported model, and has the capability to visualize and export the results. The attack definition module (ADM) interface is a tool to create rules and pre/postconditions for the fact-rule logic system used by SONARR. It also includes the capability to create actions that rules can trigger. The target representation development module (TRDM) interface is used to create the other objects required for model networks. This includes entity representation through containers and links, the facts and common properties that are used for the SONARR fact-rule logic, and supplemental custom property objects.

In addition to objects that are defined by system users as part of the model and attack technique rules, some objects are created during traversal processing. These objects are reality paths, connections, and variants. They are important to the functionality of the SONARR algorithm and its ability to find multiple unique pathways through a network without the pathways affecting each other through fact alterations from rules triggering.

The remainder of this section provides a detailed explanation of the objects used by SONARR. It also discusses the logic used for the traversal of a network, and how the inclusion of visualization helps users understand SONARR processes and results.

### 3.1. Network Objects

This section explains SONARR's network objects, at a high level. It discusses what they are and how they interact to create a model network. First, containers and links are discussed. Then, following this, facts, common properties, custom properties, and rules are discussed.

### 3.1.1. Containers and Links

The first objects covered in this section are the entities of the network: the containers and links. These objects are 'entities' due to their role in representing components within the network that can have their fact states altered by rules. The container object can represent a computer, an operating system, a program, or a network device like a firewall or a router. Containers hold facts that indicate the state of various attributes of the container. Rules use these fact states to determine if the container satisfies their trigger conditions, which can lead to the alteration of other facts within the network. Figure 1 depicts a container.

Containers can be logically nested. To accomplish this, a parent container can be specified. This, at creation, automatically generates links between the new container, the parent container, and all other containers that have the same specified parent.

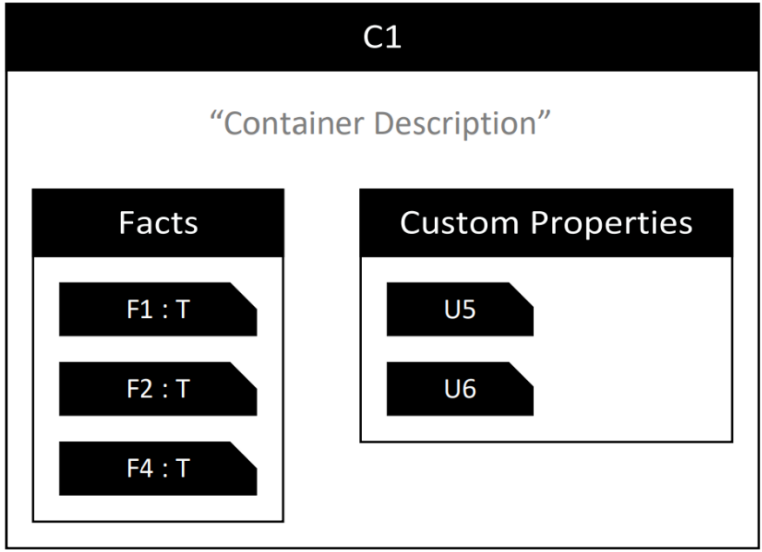

Figure 1. Simple container.

Links interconnect containers in SONARR. They represent physical device interconnectivity or other paths of influence. There are many different types of links between network components that can be represented by SONARR, ranging from ethernet wired connections to SSH tunnels to physical USB stick-based data movement. Like the container, the facts within the link entity indicate the link's current state. These facts can represent damage to a cable that causes data loss during transmission, or a wireless connection that is interfered with or blocked, reducing data thruput or disconnecting users entirely. Links' fact states also play a key role in network traversal processing. Rules evaluate fact states to see if their preconditions are met to allow them to be executed. Figure 2 depicts a link, including its facts and custom properties.

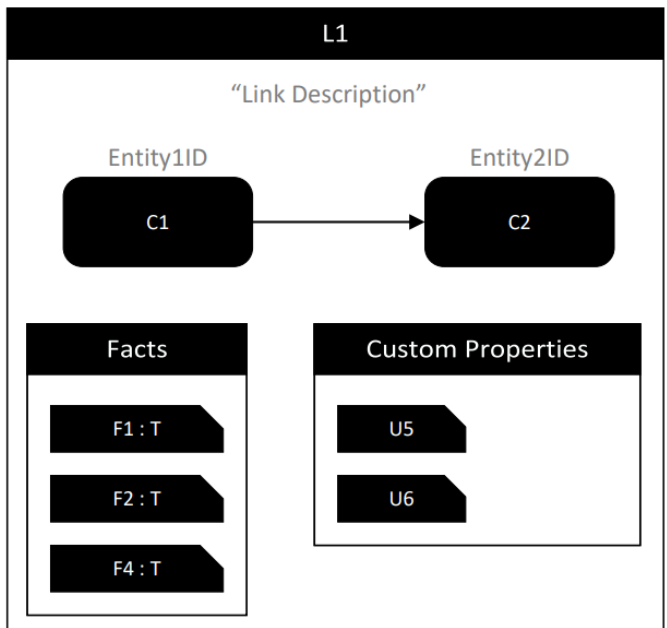

Figure 2. Simple link.

### 3.1.2. Facts

A SONARR network is built using containers and links. Both of these, as was previously discussed, will typically contain facts. Thus, the next object type to discuss is the fact. Facts store a binary value for each attribute of each entity in the network. They are not shared between entities. Common properties are used to associate facts from different objects that relate to the same thing. For example, an entity could have a fact associated with a user privilege, and another entity could also have a fact with this same association. However, this will not be the same fact used in the first entity. The facts will have different ID values and be different logical objects, allowing for their values to change independently. This individuality is essential to represent real-world entity states. These states create the unique scenarios that SONARR is built to assess. Figure 3 depicts a fact.

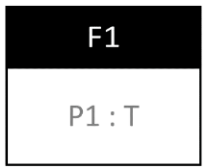

Figure 3. Simple fact.

### 3.1.3. Common Properties

Because facts are unique to their entities, a mechanism is needed to associate facts that, within the fact-rule logic, store the same type of information for any entity with that fact type. Another object, the common property, was developed for this purpose. A given common property is associated with facts that store the same type of information, effectively relating them by the common property's description. For example, two facts can be in different entities, one with a true value and the other with a value of false, and both facts can represent the user privilege attribute of the entity. These facts represent the same thing, so they are associated with the common property that has the description of "user privilege". The fact-rule logic treats the two facts as unique, but of the same type. Figure 4 shows the relationship between facts, common properties, and custom properties.

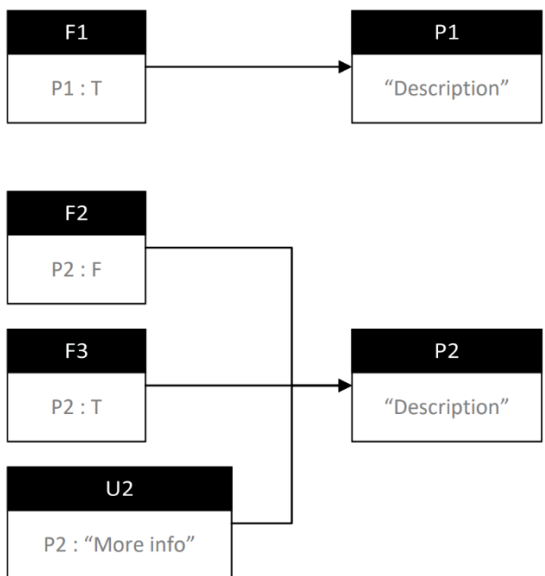

Figure 4. Relationship between facts, common properties, and custom properties.

3.1.4. Custom Properties

Custom properties are objects that can be associated with a common property and attached to a fact and/or an entity. They store string values. These are currently used as a supplemental description mechanism, but it is planned that these will become part of the fact-rule logic system in the future. Currently, custom properties are not processed by the SONARR algorithm, and they only exist to provide additional information on the front end.

3.2. Rules

Now that links, containers, facts, and common properties have been discussed, the next type of object to be introduced is rules, which define the SONARR system's logic. There are two types of rules used in SONARR: normal rules and generic rules.

A normal rule assesses the values of specific facts that are defined when the rule is created. The values of the specified facts in the network must match the values of the preconditions of the rule for it to trigger. Normal rules set the value of specific predefined facts (which can be associated with common properties) within the network if they are triggered and run. SONARR normal rules provide the same function as traditional rules in fact-rule expert systems and the Blackboard Architecture.

The second type of rule is the generic rule. Generic rules use common properties for their pre- and postconditions. Both generic rules and common properties were introduced in [1]. Generic rules are assessed in terms of three entities involved in a connection: a first container, a link, and a second container. When SONARR is assessing a connection, generic rules look at these three entities, and determine if the preconditions match the facts present in the connection. If all facts and their values match, the postconditions specified by the rule are applied to the entities. Common properties allow a rule to be assessed across any connection in the network, providing common (generic) functionality to all link-connected containers.

Figure 5 depicts rule conditions. In this figure, RC1 stores a fact and a value, and RC2 stores a common property and a value. Figures 6 and 7 show generic rules and normal rules, respectively.

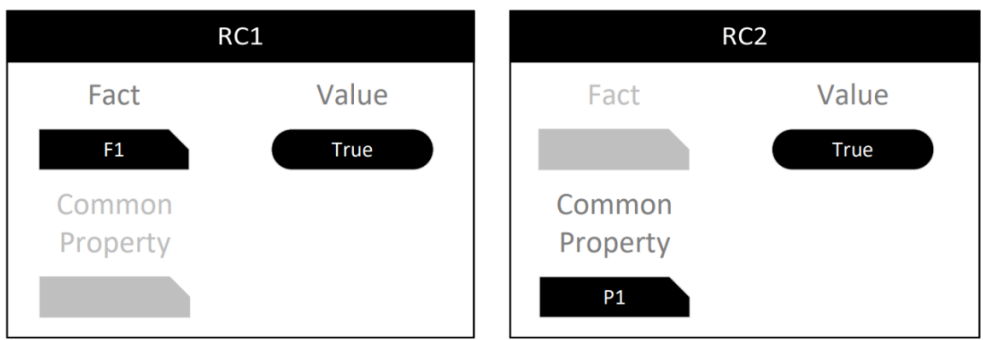

Figure 5. Rule conditions that differ in rule/fact logic properties.

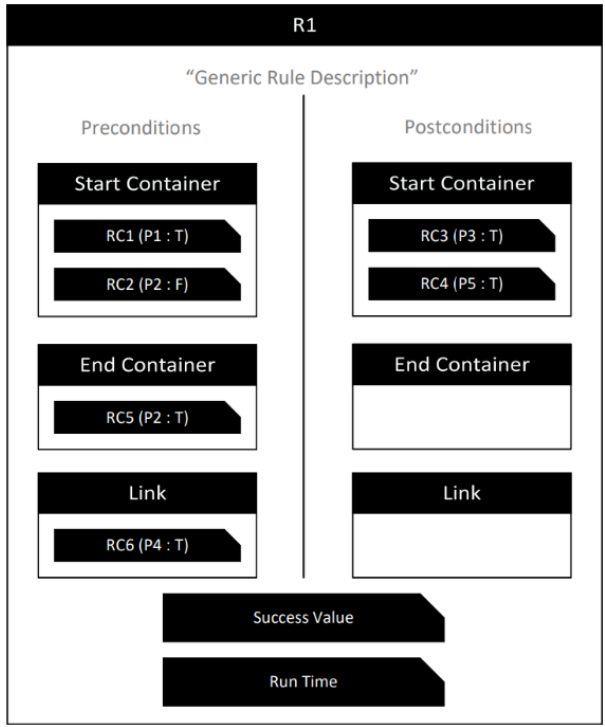

Figure 6. Depiction of a generic rule.

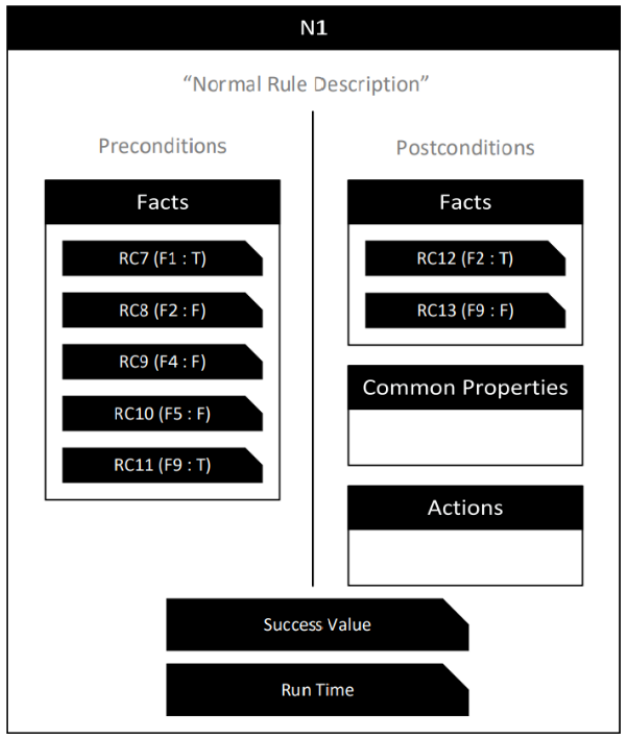

Figure 7. Depiction of a normal rule.

### 3.3. Variants, Connections and Reality Paths

Several objects are used by the SONARR algorithm during traversal processing. These objects are variants, connections, and reality paths. They are discussed in the following subsections.

### 3.3.1. Variants

Variants are in-memory versions of entities that hold different fact values than the entity in the original network. They are stored in two locations: a global collection of variants, and an 'active variant' collection within each reality path object. Containers and links are separated into type-specific lists at each of these locations.

The first location, the global collection, allows variants that are created in one reality path to be utilized as variants in other reality paths. Any variants created during SONARR processing that have a configuration of facts that does not already exist are added to this collection. These variants are then referenced by any paths that create the same fact configuration, saving memory space by preventing the storage of unnecessary variant duplicates.

The second location variants are stored is the active variant lists which store the most recent entity variants created within a reality path. Only one instance of an entity is included in the list at a time. This means that, if a container variant already exists in the list, it will be replaced with a new variant if the traversal reaches and modifies that container again during traversal processing. This allows the reality path to retain the current state of each entity altered, effectively creating an in-memory network for each reality path that is always up-to-date.

Figure 8 depicts how variant entities reference the unaltered facts of base entities, while storing facts whose values have changed via rules running.

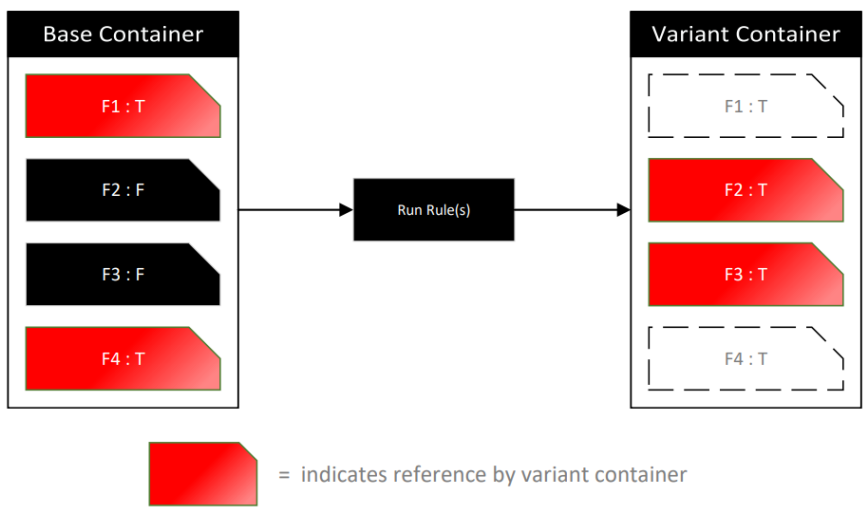

Figure 6. Depiction of a variant container referencing its base for facts that have not been altered.

### 3.3.2. Connections

Connection objects are created during traversal to store each container-link-container connection that the algorithm traverses across. The entities within this object start off as in-memory clones of the current state of the relevant entities, either from the original network or from the active variants that match the identifiers of the entities involved.

During the rule-running process, facts within entities can change, and those changes cannot be made to the original network, since all reality paths use the same original network. The changes cannot be directly applied to the associated active variant, because active variants are references to an object in the global variant list. These cannot be altered, as multiple connections may reference the same variants, and changing facts in a variant stored in the global list would alter the entities in other reality paths. Cloning the entities and performing the rule-running process in memory keeps the original network and variant fact configurations accurate. After postconditions are applied to an entity, the active variants list is updated to reference a variant matching that entity's fact configuration. If no variant in the global list exists for that configuration, then a new variant is created and added to both the global and active lists. Figure 9 depicts a connection object.

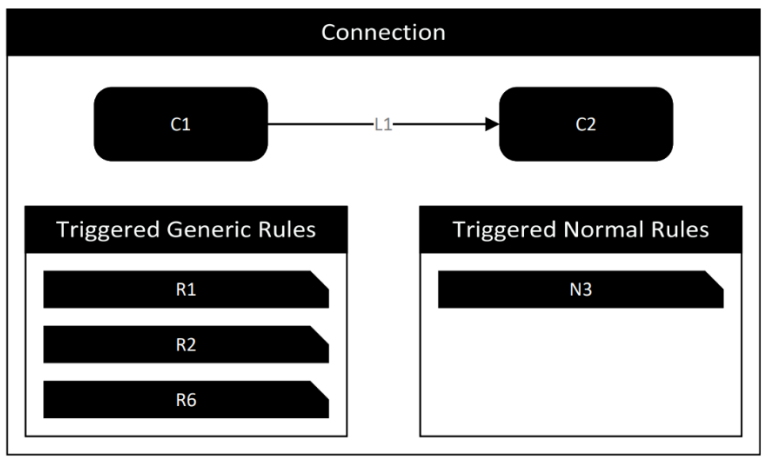

Figure 7. Depiction of a connection.

### 3.3.3. Reality Paths

The last object type is the reality path. This object stores a unique path taken through the network, including tracking the order the path takes. A reality path also tracks the history of entity fact states, the rules that are triggered at each link, the environmental fact states specific to the reality path, and the active variants of the specific reality path. All of these objects, except for environment facts, were described in previous sections. The environmental facts associated with reality paths are facts that are not associated with a common property. They are unique to each reality path, and they can only be read or modified via normal rules. They are accessible by rules that relate to any entity, and to the network as a whole. Figure 10 depicts a reality path object.

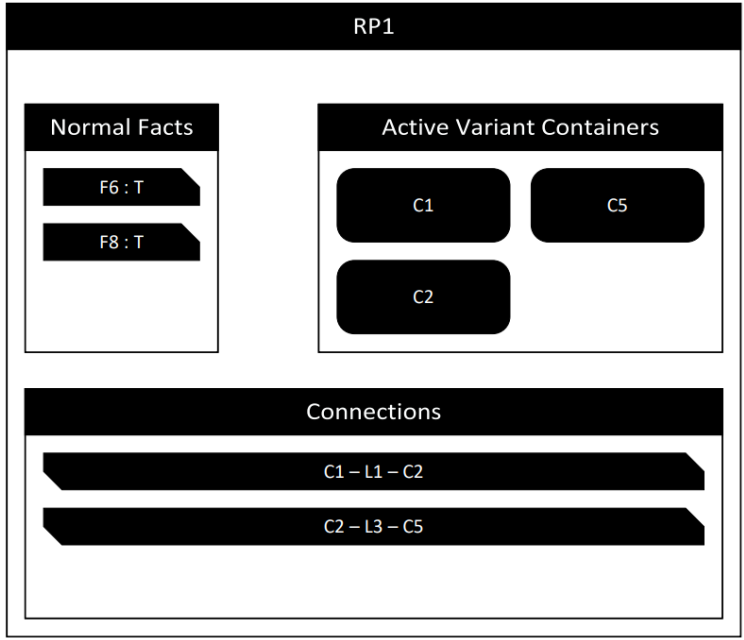

Figure 8. Depiction of a reality path.

## 4. SONARR Traversal Algorithm

SONARR's traversal algorithm is conceptually based on the basic depth-first search. It identifies paths through the network and stores them in reality paths for assessment. The algorithm creates its first reality path from the user-defined start container and, as it traverses the network, branches (i.e., creates new) reality paths at the outgoing links of each node it reaches. The reality paths are built by cloning the current path, including its active variants and the connections up to that point, and adding the next connection under rule evaluation to the path. Once the new connections are added, the reality path is placed in a processing list to be continued later and the cycle repeats until there are no more reality paths to be processed. Figure 11 presents a flowchart of the traversal algorithm.

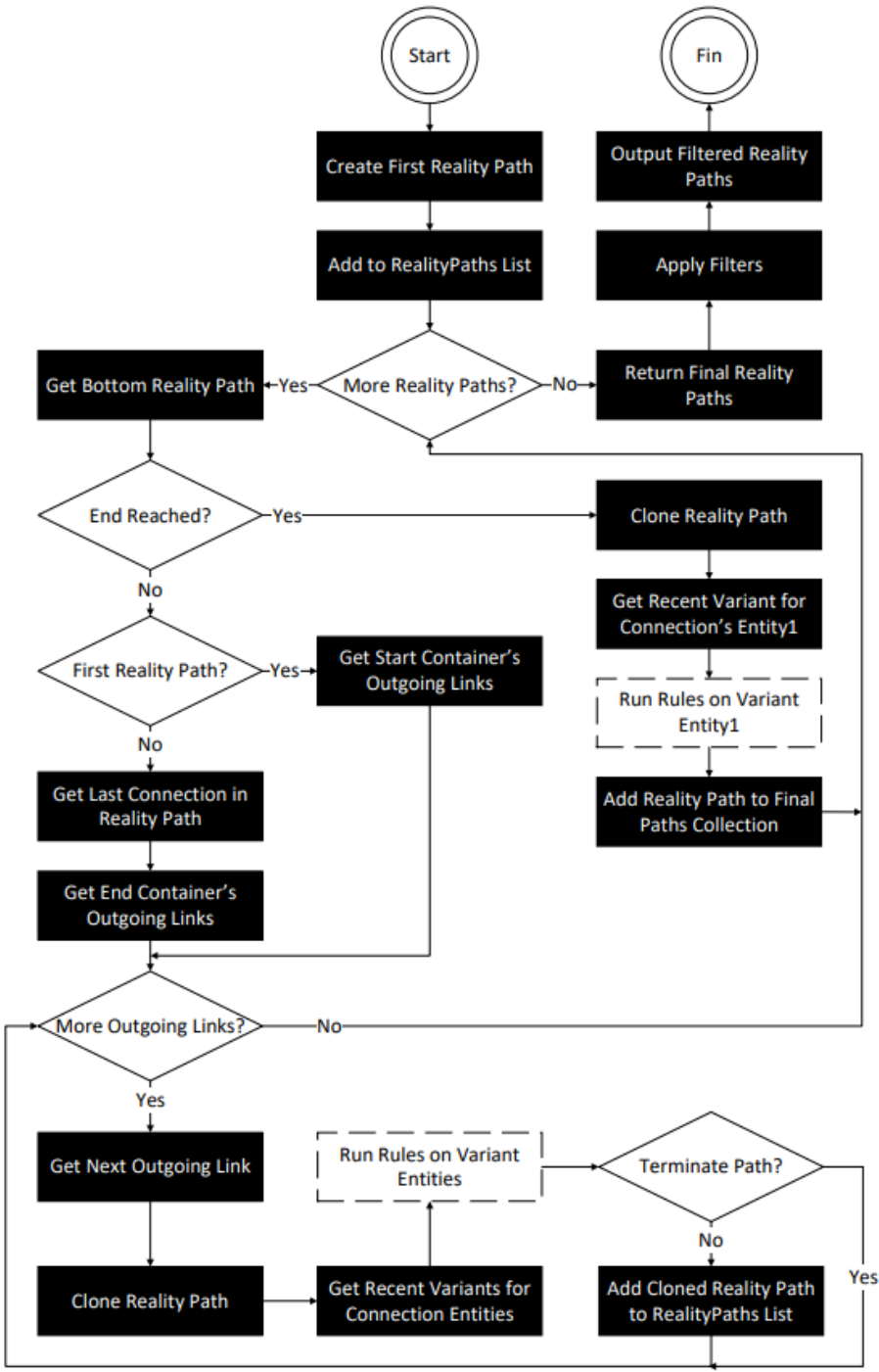

Figure 11. Flowchart of the SONARR traversal algorithm.

### 4.1 Traversal Simplification Methods

Two methods were implemented to expedite the traversal process. The first method applies a limit to the number of generic rules which can trigger for each connection. Second, a path looping check was developed that helps prevent unnecessary repeat traversals across connections.

The first method, the rule-running heuristic, uses a user-defined integer value within the range of 1 to 100. It is used during rule assessment for each connection during the traversal process. When assessing a connection, the rules are evaluated and triggered. During this process, the number of triggered generic rules, which is recorded in the connection object, is compared to the rule-running heuristic limit value. If the heuristic value has been reached, no more rules are assessed on that connection but the reality path does not necessarily terminate. The changes already made still apply and reality path traversal could continue on through the network. Figure 12 depicts how the rule-running heuristic applies to the triggered generic rules. It does not limit the number of triggered normal rules.

Rules contain a decimal success value, representing the percent chance of the rule successfully triggering. This value is user-defined at rule creation and does not change. All rules are ordered by their success value before running SONARR. The ordering of rules prioritizes the evaluating of more effective rules first. It is a mitigating measure against the risk of effective generic rules being ignored due to the rule-running heuristic.

This method of traversal simplification reduces the amount of time spent on each connection when traversing a network with a large number of generic rules. It can also create longer reality paths, if a path loops back to a connection and previously skipped rules are triggered.

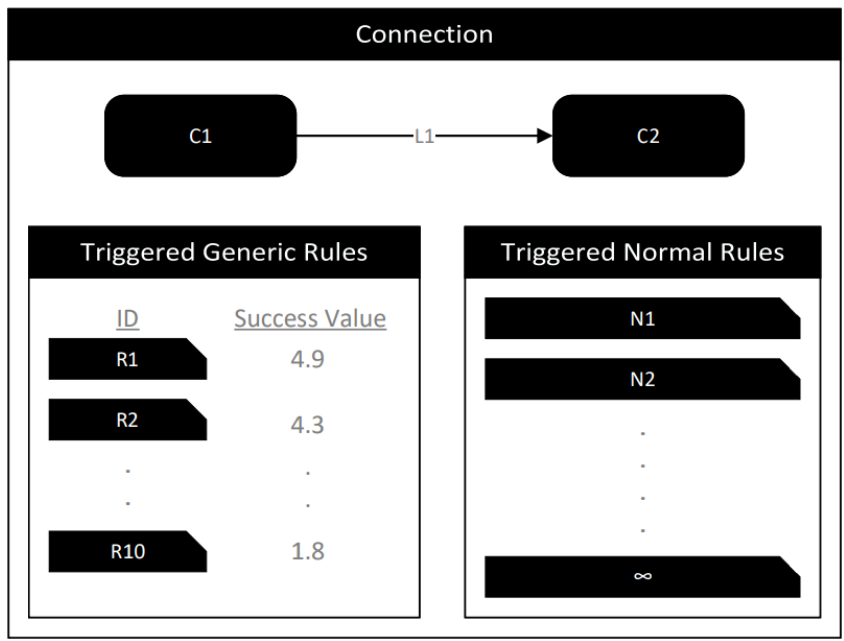

Figure 12. A connection with ten triggered generic rules.

The second method, shown in Figure 13, checks for potential unnecessary looping. It is called the path termination heuristic. Previous versions of SONARR had a 'link cap' that prevented traversal across a link if it has already been traversed a given number of times. This could prevent unnecessary repeat traversals; however, it could also prevent repeat traversals [34] required for some valid paths. The path termination heuristic addresses this by assessing rules at each connection, as usual, and then also checking that connection's entities, after the rule runs, against all previous connections and their entities in the reality path. If an exact match is found, meaning the exact state has already been explored previously, the reality path is terminated. All of the facts in all three connection entities must be the same as the previously traversed connection for the path to be terminated. If one or more facts have different values, the connection is considered new and the path traversal continues.

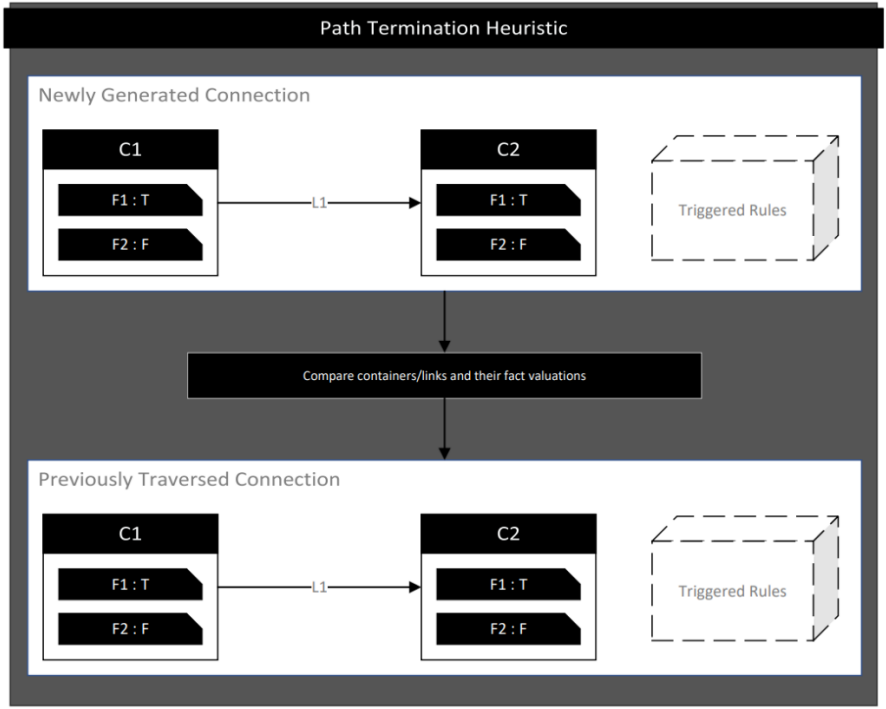

Figure 93. Path termination heuristic evaluating two matching connections.

**5. Turning Real-World Networks into SONARR Networks**

This section describes how SONARR objects are used to represent real-world networks. To do this, a scenario is presented.

The scenario is a simple network that is loosely based off of configurations used by small businesses and corporate work groups. In this network, which is shown in Figure 14, traffic flows to and from the internet through a firewall. A router is connected to this firewall. The router is also connected to two switches, a computer workstation, and a database server. One switch has four workstations connected to it, and the other switch has two workstations connected to it. One workstation in the group of four is connected directly to one workstation in the group of two, via a logical SSH connection. All of the connections are paths via which data can travel.

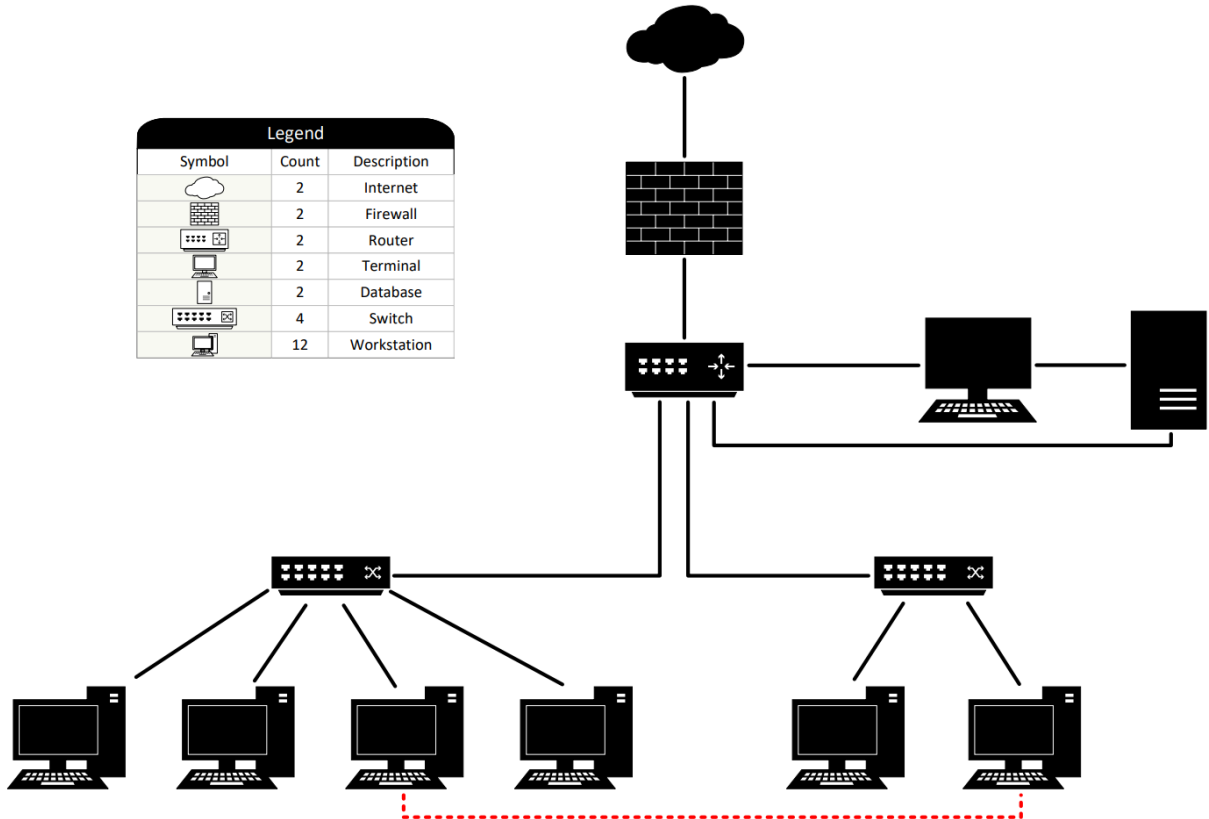

Figure 14. Basic network diagram.

Implementing a model of a real-world network in SONARR involves building a representation using SONARR's objects. When creating a SONARR model, the relationships between network objects must be evaluated. The model builder must consider which aspects of the network architecture are relevant to what an attacker could utilize to exploit the system (i.e., which devices to model and which of their properties are relevant).

Each node in the network is represented as a container object. Data-traversal connections are represented with links. A link is needed for data transit in both directions. Each container can be populated with facts representing information about the particular device it represents. Real-world networks cannot be modeled with every possible detail. Thus, each container must have sufficient detail in its facts so that all information relevant to any rules (representing exploits) is present so that they can could be triggered, if applicable, during traversal.

To accomplish this, fact configurations in containers and links must be created in a standardized way. Likewise, rules must be populated with condition components in a standard way that matches models' fact configurations. By using a rigorous and standardized approach to creating and configuring SONARR model objects, SONARR can be used to find areas of vulnerability that may not be readily identified otherwise.

## 6. Experimentation

Three model networks have been created to demonstrate and assess the efficacy of SONARR. These three networks range from a relatively simple to complex.

The first model is a small network. It is designed to make it easy to understand how the fact-rule logic works. The second model approximates a network that might be used in a sophisticated home office or very small business. It has rules that loosely model a firewall attack and file transfer. The third model is a larger, more complex network that includes only the rules and facts needed to let the traversal algorithm simulate movement throughout the network. The goal of model 3 is to facilitate analysis of how different numbers of connections to and from containers and the overall size of a network affect the reality path generation process.

The tests described in this section were run on a computer with an AMD Ryzen 7 5700U CPU with 8 GB of RAM.

**6.1. Model 1**

The first model has a simple collection of containers and links. It consists of 5 containers, 10 links, 20 facts, 10 common properties, no custom properties, 9 generic rules, no normal rules, and no actions. It is shown in Figure 15.

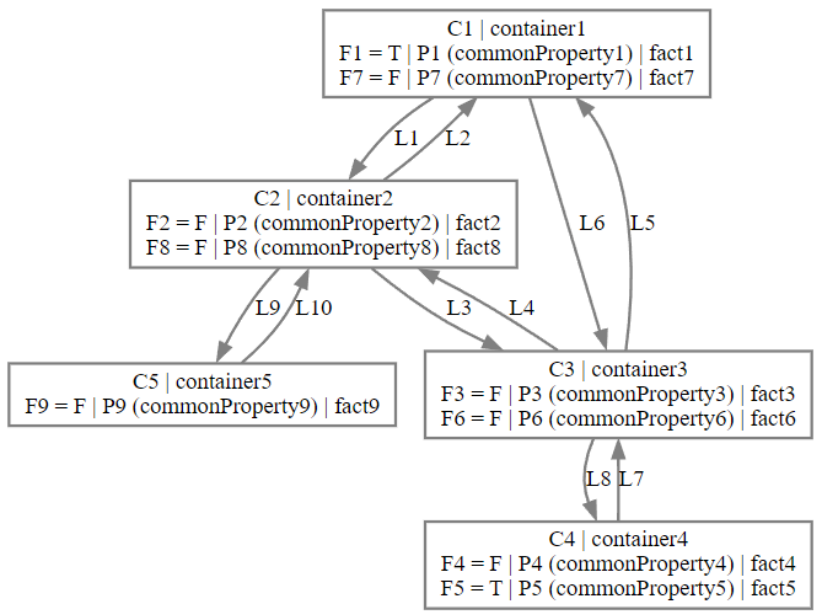

Figure 10. Model one visualization from SONARR.

Along with the facts shown in Figure 15, each link also holds a fact, which is associated with P10, that indicates the open/closed status of the link.

This model facilitates the demonstration and assessment of SONARR's performance under simple scenarios. The generic rules create a chaining effect from the start container to the end container and the scenario is designed with a simple goal to reach. In this model, the links can close after a traversal across them, to allow analysis of how that affects the results. The effect of different start and end container locations within the network can also be analyzed and compared to the other scenario results. Table 1 lists, in a simplified format, the generic rules implemented in the first model. Rules 1 to 7 have a chaining configuration that requires containers to be visited in a particular order.

Table 1. Simplified versions of the model 1 generic rules.

| ID | PREStart | POSTStart | PREEnd | POSTEnd | PRELink | POSTLink |
|---|---|---|---|---|---|---|
| 1 | P1:T, P7:F |  | P2:F, P8:F | P2:T | P10:T |  |
| 2 | P2:T, P8:F |  | P3:F, P6:F | P3:T | P10:T |  |
| 3 | P3:T, P6:F |  | P4:F, P5:T | P4:T | P10:T |  |
| 4 | P4:T, P5:T |  | P3:T, P6:F | P6:T | P10:T |  |
| 5 | P3:T, P6:T |  | P1:T, P7:F | P7:T | P10:T |  |
| 6 | P1:T, P7:T |  | P2:T, P8:F | P8:T | P10:T |  |
| 7 | P2:T, P8:T |  | P9:F | P9:T | P10:T |  |
| 8 |  |  |  |  | P10:T | P10:T |
| 9 |  |  |  |  | P10:T | P10:F |

The rules in model one are designed to be triggered in a specific order through the network. The last two rules affect the general traversal of the network. Rule 8 is the general traversal rule that triggers on every connection that has P10 set to true, effectively allowing the traversal of every 'open' link. Rule 9 is a traversal rule that closes the link after crossing, preventing repeat traversals of that link.

### 6.2. Model 2

The second model is a simplified implementation of a prototypical real-world network. It consists of a firewall, a router, two switches, five workstations, one terminal, and one database server. The containers are populated with facts that indicate the basic traversability, administrator privileges, and exploits triggered by attackers. The model consists of 12 containers, 26 links, 55 facts, 7 common properties, no custom properties, 6 generic rules, 5 normal rules, and no actions. It is shown in Figure 16.

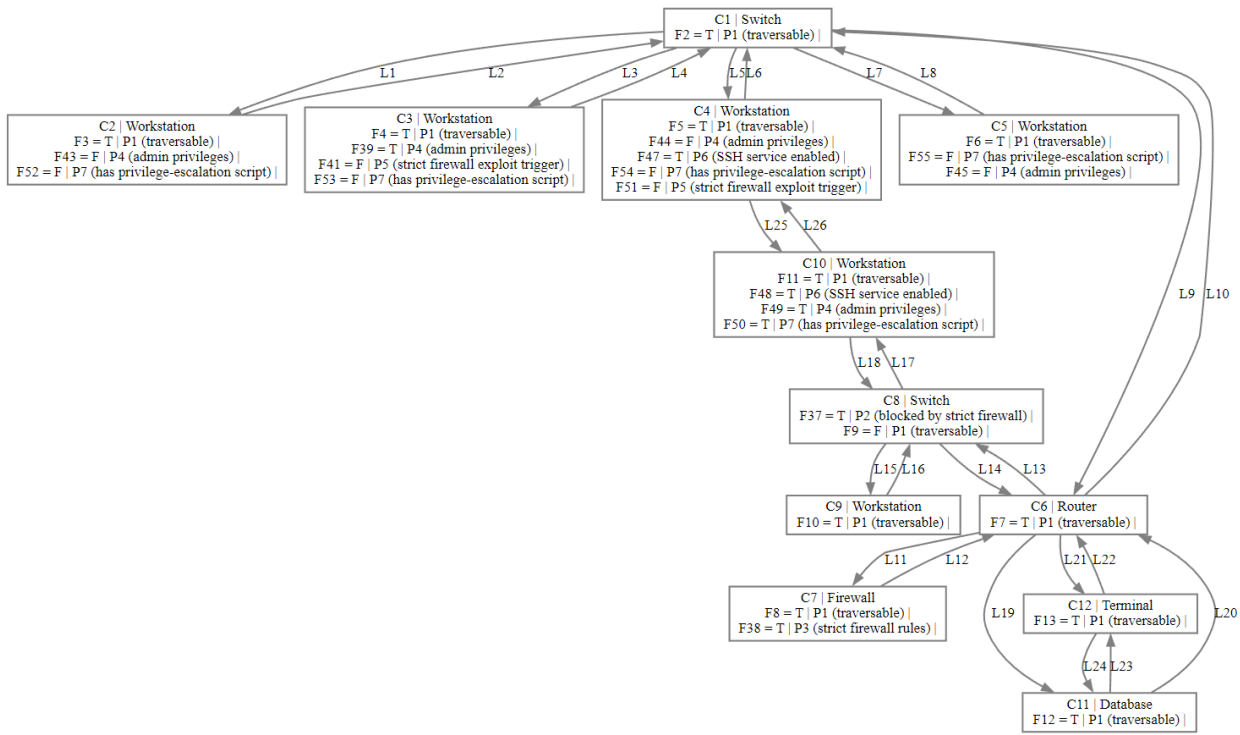

Figure 11. Model two visualization from SONARR.

This model was created to demonstrate and assess how facts in one section of a network can affect facts in another section of a network. This is demonstrated, for example, when traversing from start containers C2, C4, or C5 to end container C9. Container C9 is only connected to C8, and C8 is blocked from outside access by firewall rules. To reach container C9, these firewall rules must be disabled. In this model, this is achieved by setting C8's fact F37 to false. When SONARR is run on this model, it is shown that the only way for this to occur is for C3 to be traversed through. Container C3's combination of the admin privileges and the strict firewall exploit trigger facts will initiate a chain of generic and normal rules that allow a traversal rule to run on a subsequent connection into C8, allowing access to C9.

This model also shows how limitations an attacker might face can be identified by SONARR. For example, C4 has a strict firewall exploit trigger fact, just like C3. However, C4 cannot disable the strict firewall rules as its admin privileges fact is set to false. Container C4 has an enabled SSH service and C10 also has an enabled SSH service and a privilege-escalation script. While some reality paths may involve escalating privilege on C4, every reality path from a start container of C2, C4, or C5 to an end container of C8 will go through C3 first, as access to C10 can only be achieved by traversing C8. Because of this, C4 is not a useful node for disabling the firewall, as C4 cannot unblock C8 until after C8 has already been unblocked. Tables 2 and 3 list the rules for model two.

Table 2. Simplified versions of the model two generic rules.

| ID | PREStart | POSTStart | PREEnd | POSTEnd | PRELink | POSTLink |
|---|---|---|---|---|---|---|
| 1 | P1:T | | P1:T | | P1:T | P1:T |
| 2 | P1:T | | P1:F, P2:F | P1:T | P1:T | |
| 3 | P4:T, P5:F | P5:T | | | | |
| 4 | P6:T, P4:T, P1:T | | P6:T, P1:T | | P1:F | P1:T |
| 5 | P6:T, P7:T, P1:T | | P6:T, P7:F, P1:T | P7:T | P1:T | |
| 6 | P7:T, P4:F | P4:T | | | | |
| 7 | P1:T | | P1:T | | P1:T | P1:F |

Table 3. Simplified versions of the model two normal rules.

| ID | PREConditions | POSTConditions |
|---|---|---|
| 1 | F38:T | P2:T |
| 2 | F38:F | P2:F |
| 3 | F40:F, F41:T | F40:T |
| 4 | F40:T, F38:T | F38:F |
| 5 | F51:T, F40:F | F40:T |

**6.3. Model 3**

The third model is a more complex, arbitrary expansion of the first model. It consists of 33 containers, 78 links, 78 facts, one common property, no custom properties, two generic rules, no normal rules, and no actions. The only facts that this model contains are the link facts required for triggering the traversal generic rule.

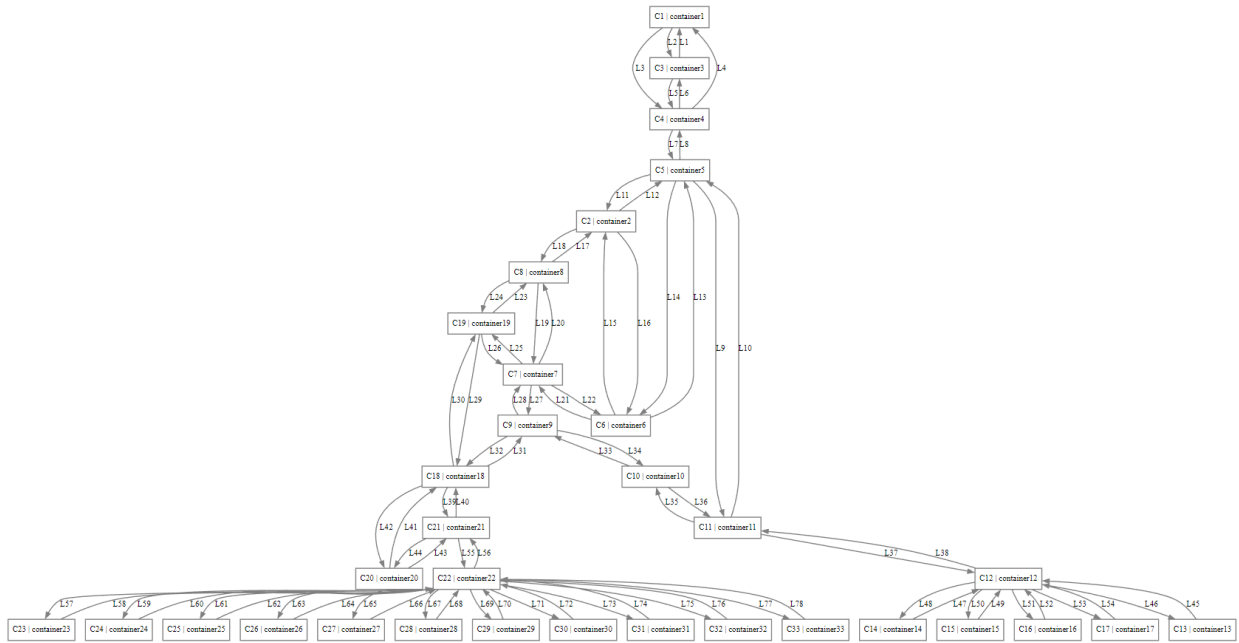

Figure 12. Model three visualization from SONARR.

The containers and links are created as depicted in Figure 17. One fact is included in each link which is associated with P1 and indicates the open or closed status of the link. Table 4 lists the rules used in model three.

Table 4. Simplified versions of the model three generic rules.

| ID | PREStart | POSTStart | PREEnd | POSTEnd | PRELink | POSTLink |
|---|---|---|---|---|---|---|
| 1 |  |  |  |  | P1:T | P1:T |
| 2 |  |  |  |  | P1:T | P1:F |

The purpose of this model is to allow further analysis of how varying styles of network architecture affect reality path generation and traversal by facilitating the development of scenarios which close off links in key locations of the network. Models 1 and 2 allow the analysis of small and medium sized networks; however, real-world networks are often larger and more complex. Thus, assessing SONARR reality path creation and traversal on a larger model helps characterize its comparative strengths and weaknesses, relative to network size.

## 7. Results Analysis

This section presents the data collected from analyzing each model in SONARR and analysis of the SONARR system's performance. Each model was created as described in the previous section and then it was analyzed in SONARR. First, the debug build was used to monitor metrics like the number of connections created and the memory usage. The release build was used to collect time metric data. Once SONARR finished its traversal – either by being stopped manually via the front end's stop button or automatically with timers set for testing purposes – the results were analyzed. Analysis was done manually using the final reality path selection capabilities, the history of entity fact states within connections, and dynamic visualization capability. The results were juxtaposed with knowledge of what rule sets are supposed to trigger, where facts are supposed to be altered, and what paths can be traversed within the model.

The results from the three models are now discussed in the following subsections. Each subsection has a table describing the scenarios used to evaluate SONARR with the model. These tables include the user-defined start container, the user-defined end container, the network configuration settings specific to the scenario, a simple description, the fastest completion time of the scenario, the total number of reality paths found, the number of reality paths found that achieve the model's goal, the total connection count of all found reality paths for the scenario, the longest reality path length (with the number of reality paths recording that length in parentheses), the shortest reality path length (with the number of reality paths recording that length in the parentheses), the total number of variant containers created during the scenario session, the total number of variant links created during the scenario session, and the estimated memory usage of the scenario session.

**7.1. Model 1**

This section discusses the results of the SONARR sessions for the model 1 scenarios. The data for this model is presented in Table 5.

Table 5. Scenario details and results for model one.

| Scenario # | 1 | 2 | 3 | 4 | 5 |
|---|---|---|---|---|---|
| Start Container | C1 | C1 | C1 | C1 | C5 |
| End Container | C5 | C5 | C5 | C3 | C4 |
| Configuration | Omit R8 and R9 | Omit R8 | Omit R9 | Omit R9 | Omit R9 |
| Description | Only traverse with chaining rules | Traverse + close links | Traverse | Traverse | Traverse |
| Fastest Completion Time | Immediate | Immediate | Immediate | Immediate | Immediate |
| Final Reality Paths | 1 | 20 | 360 | 5 | 78 |
| # Goal Achieving Paths | 1 | 0 | 272 | 0 | 0 |
| Total Connections | 8 | 140 | 5522 | 20 | 865 |
| Longest Path | 8 (1) | 9 (6) | 22 (6) | 6 (1) | 15 (8) |
| Shortest Path | 8 (1) | 3 (1) | 3 (1) | 2 (1) | 5 (1) |
| Variant Containers Created | 8 | 10 | 12 | 5 | 8 |
| Variant Links Created | 6 | 14 | 9 | 6 | 9 |
| Memory Usage | 23 MB | 23 MB | 24 MB | 23 MB | 24 MB |

The goal for the first model is to start at C1 and find all paths to C5 that flip the value of F9 from false to true by triggering all seven generic rules that are necessary to do so. Succeeding in this goal is akin to an attacker following a path through a network to ultimately reach an area that is not directly accessible.

The first scenario for this model demonstrates the use of rule chaining as the sole method of crossing a link. Because of the path termination heuristic, the rules create a single realistic path from C1 to C5. A generic rule alters facts at every connection for this model. SONARR evaluates crossing every connection from every container in the model, but only one connection from each container produces a valid reality path. This scenario completed nearly instantaneously and resulted in a single reality path with eight connections being identified. During this test eight variant containers were created, along with six variant links. This scenario achieved the goal of altering the F9 valuation and it did so by triggering the rules consecutively.

The second scenario for this model demonstrates the use of rule chaining while closing links during crossing, effectively preventing repeat traversal over that link. This model uses closed links at every connection as a means of shortening the number of assessable links. A real world parallel to this is a piece of malware that sabotages connections to the network. Closing a link, in this model, involves triggering R9 which then flips a fact in the current connection's link from true to false. This prevents other rules from triggering because the preconditions requiring that fact be set to true are no longer satisfied. The scenario also completed nearly instantaneously. It found 20 reality paths, with 140 total connections. The six longest reality paths recorded nine connections from the start container to the end container, and one reality path recorded the shortest length of three connections. A total of ten variant containers were created as the reality paths were generated, as well as 14 variant links. None of the reality paths were able to achieve the goal of flipping F9 to true, even though 20 paths reached the final container.

The third scenario also demonstrates the use of rule chaining. However, it does not close links when they are traversed. Instead it allows the path termination heuristic to stop any looping reality paths. Scenario three was also able to complete processing nearly instantaneously. The traversal processing produced 360 reality paths, because paths could re-cross links, if the path termination heuristic allowed it. With this, 5,522 total connections were created. The longest path length was 22 connections. There were six reality paths of that length. The shortest was still three connections. There was only one reality path of that length. There were a total of 12 variant containers and nine variant links. Because this scenario did not close links after crossing them, allowing repeat link traversals, SONARR was able to find 272 unique reality paths that triggered all 7 rules necessary to satisfy the goal. One of the paths matched the path created for scenario one.

The last two scenarios demonstrate the impact that the start and end container locations have on SONARR results, based on the layout of the model network. Like the third scenario, both of these scenarios have rules 1 to 7 active and keep links open after crossing them.

The fourth scenario seeks to find paths from C1 to C3 which, based on the model's architecture, prevents SONARR from reaching C4 because the algorithm will not past the specified end container of C3. This also prevents over half the rules from triggering, because they all require the algorithm to first reach C4 and trigger rules associated with facts in that container in order to have their preconditions satisfied. Not reaching C4 is the correct result because the selected end container is C3. All of the paths achieve this, based on the network architecture.

Scenario four completed nearly instantaneously. It resulted in five reality paths and 20 total connections. The longest path had six connections. The shortest path had two connections. In both cases, only a single path of this size was produced. Five variant containers and six links were created.

Scenario five also completed nearly instantaneously. It provided 78 reality paths with a total of 865 connections. Eight paths that had the longest number of connections (15) and one path had the shortest number of connections (five). Eight variant containers and nine links were created. As expected, none of these paths were able to flip F9 to true because the specified end container is in the middle of the generic rule chain, preventing the rule chain from completing the overall goal. Any path that reaches the end container is final, so the reality paths stop and cannot continue to the container holding F9, even though the specified end container is pivotal to achieve the overall goal of flipping F9 to true. This scenario demonstrates the importance of connection traversal order and the influence end container selections can have on the SONARR algorithm, based on a network's architecture.

Overall, model one used a chain of simple generic rules to demonstrate and evaluate traversals with the simple goal of flipping F9 to true. As shown in Table 3, scenario one, as expected, successfully found the reality path. Scenario two, as expected, did not find any reality paths that achieve the scenario goal, because a rule closed paths after crossing. Finally, scenario three, as expected, found multiple paths that achieve the goal, because of allowing repeat traversals. These three scenarios demonstrate the traversal functionality in SONARR and show how the results can differ based on the facts and rules implemented in a network. Scenarios four and five demonstrated, in particular, how the design of a network can influence SONARR's results, especially within a small model.

**7.2. Model 2**

This section presents and discusses the results of the SONARR sessions for the model 2 scenarios. Data for model 2 is presented in Table 6.

Table 6. Scenario details and results for model two.

| Scenario # | 1 | 2 | 3 | 4 | 5 |
|---|---|---|---|---|---|
| *Start Container* | C2 | C2 | C2 | C2 | C2 |
| *End Container* | C9 | C9 | C9 | C9 | C9 |
| *Configuration* | Omit R7 \| F39:F | Omit R7 | Omit R7 \| F44:T | Omit R7 \| F37:F, F38:F | Omit R1 |
| *Description* | C3 does not have admin privileges | Traverse with chaining rules | C4 has admin privileges | Disable strict firewall rules | Scenario 3 with link closing |
| *Fastest Completion Time* | 00:00:01 | 00:00:06 | 00:01:33 | 00:00:13 | 00:00:02 |
| *Final Reality Paths* | 0 | 2444 | 123919 | 4056 | 1144 |
| *# Goal Achieving Paths* | 0 | 2444 | 123919 | 4056 | 1144 |
| *Total Connections* | 0 | 46372 | 2872014 | 75400 | 19168 |
| *Longest Path* | 0 (0) | 25 (144) | 30 (900) | 25 (216) | 21 (108) |
| *Shortest Path* | 0 (0) | 7 (1) | 6 (1) | 5 (1) | 7 (1) |
| *Variant Containers Created* | 10 | 19 | 19 | 17 | 16 |
| *Variant Links Created* | 18 | 26 | 25 | 26 | 26 |
| *Memory Usage* | 27 MB | 35 MB | 392 MB | 399 MB | 402 MB |

The goal of the second model is to reach C9, which is inaccessible without triggering certain rules. The start and end containers of the five scenarios for this model are C2 and C9, respectively. These containers are on opposite sides of a firewall-restricted container (C8) and don't play a role in the rule-chaining within the model. Rather, connections to and from C2 and C9 only trigger the basic traversal rule.

Scenario one is a model with no traversable paths between C2 and C9. With this scenario's initial container/fact configurations, C9 is inaccessible from the starting point, due to the firewall rules blocking traversal through C8 and C3 not having the admin privileges required to disable them. This scenario finished in one second with zero final paths. This is an accurate result, based on the scenario design which makes it so that an attacker cannot gain access because of lacking the proper privilege to exploit.

Scenario two is identical to scenario one with the exception that, in scenario two, C3 now has admin privileges. When SONARR evaluates the connection C3-L4-C1, the trigger exploit if privileged generic rule triggers. This starts a rule chain that includes three normal rules for this connection and a fourth normal rule that triggers on all subsequent connections. This fourth normal rule is what sets C8's blocked by strict firewall fact to false, allowing generic rule R2 to set C8's traversable fact to true. This then allows traversal onward to C10 and C9. This scenario finished traversal processing in 10 seconds with 2,444 final paths and 46,372 total connections. There were 144 paths with the longest number of connections (25) and one path with the shortest (7). There were 19 variant containers and 26 links created.

Scenario three is identical to scenario two with the exception that, in scenario three, C4 has admin privileges and additional links. In scenario two, C3 had to be traversed in every reality path from C2 to C9. In scenario three, C4 can open up C8 using the same rule-chaining mechanism that C3 utilizes. C4 also has two additional links: one incoming and one outgoing. These both go to C10, a workstation container that was previously only accessible by traversing to it via C8. The combination of having an additional trigger point for the firewall rule-chaining and having a new route to take to the other section of the network creates significantly more possible reality paths. This scenario produces a shorter shortest path than scenario two did. It also produces one less variant link due to R4, which requires C4 having admin privileges, always triggering and changing L25's traversable fact to true. The scenario finished in two and a half minutes with 123,919 final paths and 2.8 million total connections. There were 900 paths with the longest number of connections (30) and one path with the shortest (6). A total of 19 variant containers were created along with 25 variant links.

Scenario four is the same as scenario two, except for two initial fact values that are changed. Container C8's blocked by strict firewall fact is now initially set to false and C7's strict firewall rules fact is now also set to false. This essentially disables the strict firewall rules from the beginning of the scenario, removing the need to traverse through C3 or C4 before reaching C8. This scenario has 1,612 more reality paths than scenario two, due to the lack of restrictions. It also produces the shortest possible path from C2 to C9. The scenario finished in 22 seconds with 4,056 final paths and 75,400 total connections. The largest number of connections was 25 which is found in 216 paths, and the shortest number of connections was 5 which is only found in one path. The run created 17 variant containers and 26 links.

Finally, scenario five is also similar to scenario two. However, instead of using R1 as the basic traversal rule, it uses R7. While R1 and R7 both require all three entities in a connection (origin container, link, and destination container) to have a true traversable fact, R7 changes the link's traversable fact to false after triggering. This closes off links from being traversed a second time, if there are no other generic rules that can apply on subsequent connections across the link. This scenario finished in four seconds. It identified 1,144 final paths and 19,168 total connections. There were 108 paths with the longest number of connections (21) and one path with the shortest (7). There were 16 variant containers and 26 links created.

### 7.3. Model 3

This section discusses the results of the SONARR sessions for the model 3 scenarios. Data for model 3 is presented in Table 7.

Table 7. Scenario details and results for model three.

| Scenario # | 1 | 2† | 3 | 4† | 5† |
|---|---|---|---|---|---|
| *Start Container* | C12 | C22 | C1 | C1 | C1 |
| *End Container* | C13 | C29 | C11 | C15 | C15 |
| *Configuration* | Omit R1 \| F39:F | Omit R1 \| F57:F | Omit R1 \| F38:F, F30:F, F33:F | **DEBUG BUILD** Omit R1 | **DEBUG BUILD** Omit R2 |
| *Description* | One container with 5 connections + close links | One container with 11 connections + close links | Traverse + close links | Whole network + close links | Whole network |
| *Fastest Completion Time* | Immediate | 00:03:50 | 00:00:36 | 20:00:00 | 20:00:00 |
| *Final Reality Paths* | 65 | 1200001 | 185913 | 3 | 3 |
| *# Goal Achieving Paths* | N/A | N/A | N/A | N/A | N/A |
| *Total Connections* | 522 | 24000020 | 4191202 | 48 | 48 |
| *Longest Path* | 10 (24) | 22 (441456) | 28 (8640) | 18 (1) | 18 (1) |
| *Shortest Path* | 2 (1) | 4 (1) | 4 (1) | 14 (1) | 14 (1) |
| *Variant Containers Created* | 6 | 12 | 12 | 33 | 33 |
| *Variant Links Created* | 18 | 21 | 30 | 154 | 154 |
| *Memory Usage* | 26 MB | 2.4 GB | 518 MB | 25 MB | 25 MB |

† denotes a manually stopped SONARR session.

Overall, for this model, only a minimal amount of fact alteration occurs. This happens in some scenarios where links are closed after crossing. Therefore, the importance of variant containers and links in model three is minimal and they are not discussed further. The numbers of each are still reported for consistency.

Scenarios one and two test reality path generation using the one-depth tree architectures shown in Figures 18 and 19. Each was developed to assess how SONARR session time is affected by different amounts of one-depth branched containers. The first scenario starts at C12 and ends at C13, with the link exiting the start container beginning closed. This limits the possible initial traversal space to consist solely of C12 and the 5 branching containers. This scenario completed nearly instantaneously and found 65 final paths with 522 total connections. Of these, 24 paths had the longest length of 10, and one path had the shortest length of two. This scenario shows that a model with one container and five branching containers, in the shape of a one-depth tree, results in 65 reality paths all with a unique traversal order. This is important because it gives context to the influence network architecture has on SONARR traversal processing, as demonstrated in scenario 2.

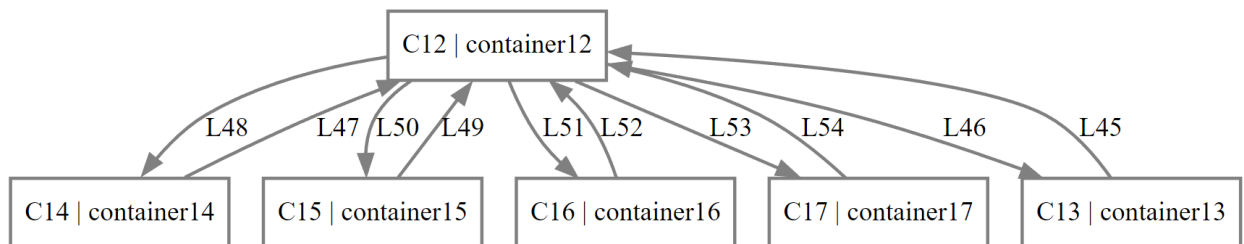

Figure 13. The one-depth tree architecture in scenario one.

Scenario two is notably different from scenario one. It has the same one-depth tree architecture as the first scenario; however, instead of five branches there are 11. This scenario was set to automatically stop if the number of paths found reached above 1.2 million. This number was chosen as a stopping point because previous testing on this scenario would exhaust available memory around 1.6 million paths in approximately 15 minutes, and would have significant trouble reaching that number. For scenario two, SONARR found 1,200,001 final reality paths consisting of 24 million total connections. This is a significant increase in both time and memory usage for a relatively small number of additional containers and links.

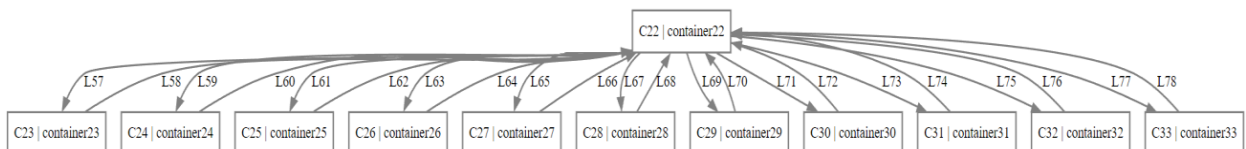

Figure 14. The one-depth tree architecture in scenario two.

The resulting number of reality paths from this type of architecture with the configuration settings described previously can be found using the equation:

$$a_n = (n-1)a_{n-1} + 1.$$

Using this formula, scenario two should identify 9,864,101 paths. SONARR did not come close to this number. It is believed that the reason for this is the large number of final reality paths and the connection objects stored inside each one. Extending the network further than this, like was done with model three, further increases the level of memory and time required, beyond the level that SONARR can presently handle.

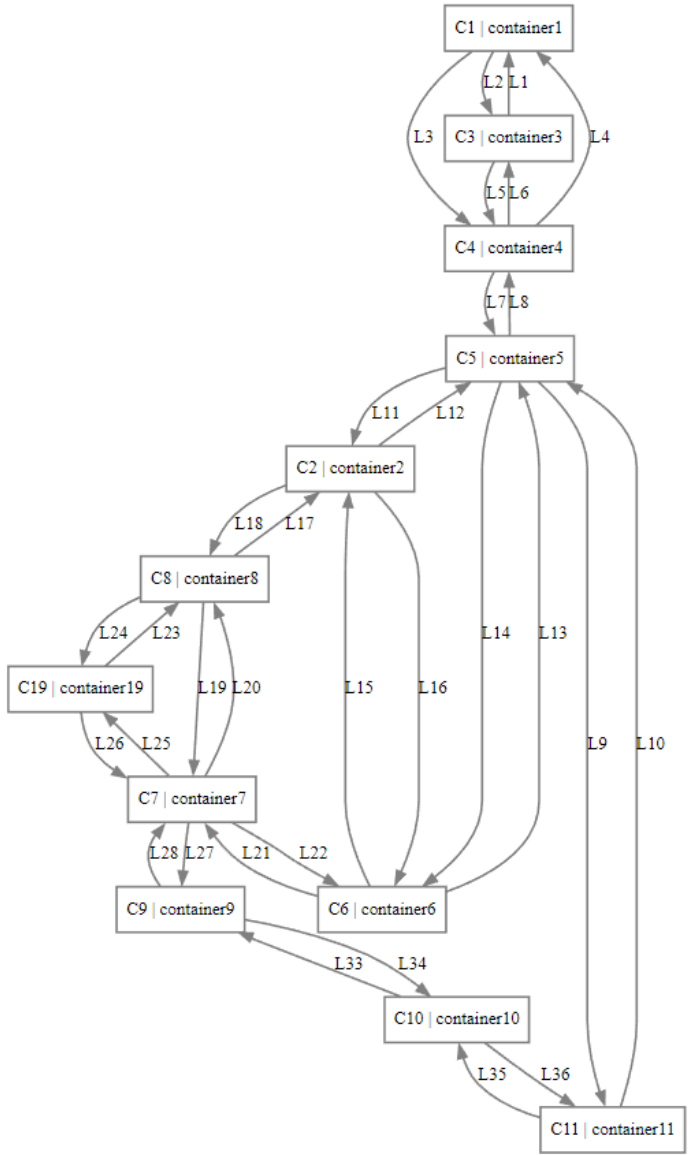

Figure 15. Section of model three used in scenario three.

Scenario three builds on scenario two and also uses 12 containers. However, this approach spreads the connections out more within the model. Because of this, even with scenario three containing more overall that SONARR needs to consider (32 links in scenario three vs. 22 links in scenario two), it was able to complete its session with 185,913 reality paths found and 4,191,202 total connections created.

Of these, 8,640 paths had the longest length of 28 connections, and one path had the shortest length of four connections. Comparing these results with scenario two indicates a strong correlation between container/link configurations of the model network and the success of SONARR traversal functionality.

Scenarios four and five test the network from one end to the other. Scenario four closes links after crossing them and scenario five leaves them open. These two scenarios were run only in the debug build for a maximum of 20 hours to see if there was change in the result metrics within this time. However, they only found three reality paths and had 48 total connections, all within the first second. Nothing else was found within 20 hours. The longest recorded path for both scenarios was 18 connections, and the shortest recorded path was 14 connections. Both tests demonstrate the computational complexity and time requirements of larger networks. This model network is large enough that, even with both the path termination and the close links heuristics operating during the traversal rule (R2) running, SONARR still takes a considerable amount of time to process the network. If allowed to run long enough, the number of reality path possibly created could exhaust device memory. Because of this, further optimization of the SONARR algorithm and the storage of results are key priorities for future work.

## 8. Conclusions and Future Work

The experimentation results have shown that the SONARR algorithm takes longer and uses more memory when traversing larger networks, which is not unexpected. The testing has also demonstrated the functionality and efficacy of the other new features and changes. SONARR now makes better use of memory for variant objects. This is a noticeable enhancement compared to previous versions. SONARR's core logical capabilities have also been expanded through the changes and additions described.

There are other areas that can be explored to enhance SONARR's performance and capabilities, such as altering the algorithm to support multi-threading. The storage approach for objects with simpler time complexities can also potentially be enhanced. More efficient code loops and referencing previously created connection objects instead of creating duplicates, much like how variants currently behave, could also provide benefit. Increasing the efficiency of processing and memory usage is key to making SONARR usable with larger networks and therefore demands attention.

Another potential area for enhancement is the addition of other filter options to better analyze the results of a SONARR session. Standard confidentiality, integrity, and availability impact scores can potentially be used to provide a glimpse of each reality path's impact on those three keys metrics for a real-world network.

The final planned area of improvement is to reassess the current heuristics, particularly the path termination heuristic, to see if they can be changed to enhance their efficiency. Currently, the heuristics prevent the continuation of a reality path if the newest connection is an exact replica of a previously traversed connection. While this does prevent unnecessary looping, as was intended when implementing it, it also could prevent some beneficial looping. Thus, a change could potentially facilitate the exploration of areas within the network that could hold different values from previously triggered rules. For example, some rules could alter facts in remote corners of a network, but the path termination heuristic could end reality paths before they can reach or use the new state of the network. Altering this behavior could increase the number of reality paths found through a network, as well as increasing the fidelity of the collection of found paths.

**Acknowledgements**

Thanks is given to Jordan Milbrath for his work on this implementation and to Anthony DeFoe for his work on visualization development for the software. Thanks is also given to other members of the project team for their feedback, testing and other contributions.